\documentclass[12pt]{article}
\usepackage{graphics,cite}

\def\Eq#1{Eq.~(\ref{#1})}

\def\beq{\begin{equation}}
\def\eeq{\end{equation}}
\def\beqa{\begin{eqnarray}}
\def\eeqa{\end{eqnarray}}
\def\bet{\begin{tabular}}
\def\eet{\end{tabular}}

\newcommand{\sect}[1]{\setcounter{equation}{0}\section{#1}}

\begin{document}

\begin{titlepage}

\begin{flushright}
CERN-PH-TH/2005-090\\
DFTT-15/2005\\
June 2005
\end{flushright}

\vspace{1.3cm}

\begin{center}
{\Large \bf Soft-gluon resummation effects\\
on parton distributions\\}
\end{center}

\vspace{5mm}

\begin{center}
{\large \bf Gennaro Corcella$^1$ and Lorenzo Magnea$^2$}\\

\vspace{5mm}

{$^1${\sl Department of Physics, CERN}\\ 
{\sl Theory Division}\\
{\sl CH-1211 Geneva 23, Switzerland}\\}

\vspace{3mm}

{$^2${\sl Dipartimento di Fisica Teorica, Universit\`a di Torino}\\
{\sl and INFN, Sezione di Torino}\\
{\sl Via P. Giuria 1, I-10125 Torino, Italy}\\}

\end{center}

\vspace{1cm}

\begin{abstract}

We gauge the impact of soft-gluon resummation on quark distributions
by performing a simple fit of Deep Inelastic Scattering structure
function data using next-to-leading order
(NLO) and next-to-leading-logarithmic (NLL)-resummed coefficient functions. We
make use of NuTeV charged-current data, as well as New Muon Collaboration
(NMC) and Bologna--CERN--Dubna--Munich--Saclay (BCDMS)\\
neutral-current results, which probe large values of $x$. Our results
suggest that the inclusion of resummation effects in global fits of
parton distributions is both feasible and desirable, in order to
achieve at large $x$ the accuracy goals of the LHC physics program.

\end{abstract}

\end{titlepage}

\sect{Introduction}
\label{intro}

A precise knowledge of parton distribution functions (PDF) in the
proton is going to be one of the cornerstones of the physics analysis
of LHC data~\cite{Giele:2002hx}, as well as a key ingredient for
studies at other high energy accelerators. In hadron collisions, in
fact, all high-$p_\perp$ final states are produced through the hard
scattering of partons, thus both potential new physics signals and
Standard Model backgrounds are affected in shape and normalization by
parton distributions.

PDF's are currently determined by several
groups~\cite{Pumplin:2002vw,Martin:2002dr,Alekhin:2002fv} through fits
to different sets of high-energy data.  All current fits are performed
at least at next-to-leading order (NLO) in perturbative QCD, while the
remarkable recent calculation of the three-loop Altarelli--Parisi
splitting functions~\cite{Vogt:2004mw,Moch:2004pa} has made it
possible to perform consistent next-to-next-to-leading order (NNLO)
fits, by restricting the data set to cross sections for which the
theoretical calculation has been performed to that order.

It is well known, on the other hand, that finite-order QCD
calculations are limited in their range of applicability by the
occurrence of large logarithms near the boundaries of phase space,
both at large and at small values of $x$. These logarithms must be
resummed, in order to enlarge the region in which perturbation theory
can be trusted. Large-$x$ logarithms, in particular, are known to be
related to soft and collinear gluon emission, and their resummation
(threshold resummation) is well
understood~\cite{Sterman:1986aj,Catani:1989ne}, and applied to a wide
range of hard QCD processes~(see, for example, \cite{Banfi:2004yd}).

In this paper we shall address the possibility of including the effect
of threshold resummations in parton fits, and we shall gauge the
impact of these effects on large-$x$ quark distributions, by
performing a simple analysis of Deep Inelastic Scattering (DIS) data.

We believe that including resummations would be useful in several
respects. From a phenomenological viewpoint, making use of resummed
predictions would allow for the inclusion of more large-$x$ data
points in parton fits. In DIS, for example, data corresponding to
values of $W^2 = Q^2 (1 - x)/x$ smaller than about $15$ GeV$^2$ are
typically excluded from the fits~\cite{Martin:1998np}, since they
cannot be accounted for by making use of NLO perturbative
results. Including resummations should lower this bound
considerably~\cite{Gardi:2002xm}. Resummations are also known to
reduce the theoretical uncertainty of QCD
predictions~\cite{Sterman:2000pu}, which would correspondingly
decrease one of the sources of error for PDF's. Finally, it should be
emphasized that, although soft-gluon resummations modify hard cross
sections only near partonic threshold, they can affect parton
distributions at smaller values of $x$ through sum rules as well as
evolution.

Taking a more formal viewpoint, including resummation effects would
bring about significant progress in the process of giving a precise
definition of a fitted {\it leading twist} PDF. Resummations, in fact,
are inevitably entangled with power corrections, which become
increasingly important near the edges of phase space. This,
however, should not be understood as an extra source of ambiguity: on
the contrary, the inclusion of resummations highlights an inherent
ambiguity, which is always present when finite-order perturbative
predictions are used to extract from data the values of operator
matrix elements.  In general, it is not consistent to attribute a
fixed twist to quantities evaluated at finite perturbative orders in a
mass-independent regularization scheme, such as dimensional
regularization. In such a scheme, one must first give a precise
definition of the perturbative contribution to all orders, which
entails a definition of power-suppressed contributions.  Such a
definition can only be given when all-order contributions have been
computed, at least in the region of phase space where power
corrections are expected to become dominant. This somewhat formal
issue could become practical when a sufficiently precise comparison
between PDF's obtained by fitting data and PDF's obtained from the
lattice becomes possible.

Finally, it should be noted that resummed predictions exist for most
of the cross sections used in global PDF fits, although with a varying
degree of accuracy. The gold-plated process remains inclusive DIS,
where, remarkably, we now have a full next-to-next-to-next-to-leading
order (NNNLO) QCD prediction~\cite{Vermaseren:2005qc}, as well as
next-to-next-to-leading logarithmic (NNLL) soft-gluon
resummation~\cite{Vogt:2000ci}, and refined QCD-motivated models of
the leading power corrections~\cite{Gardi:2002bk}; even a class of
non-logarithmic terms has been shown to
exponentiate~\cite{Eynck:2003fn}. The Drell--Yan cross section is
understood with almost the same degree of
accuracy~\cite{Vogt:2000ci,Eynck:2003fn}, with the added feature of a
recent NNLO computation of the vector boson rapidity
distribution~\cite{Anastasiou:2003yy}. This is interesting because it
has recently been shown~\cite{Alekhin20005} that reasonably
competitive parton fits can be obtained on the basis of these two
processes only. If one wishes to rely upon a wider data set,
next-to-leading logarithmic (NLL) resummed predictions exist also for the
prompt-photon production cross section~\cite{Catani:1999hs}; there,
however, phenomenological problems remain, partly associated with a
possible inconsistency of different data sets, and possibly related to
the need to perform a more refined resummation~\cite{Laenen:2000de}
and to include consistently power-suppressed
corrections~\cite{Sterman:2004yk}. Jet production in hadron collisions
is more problematic: in fact, although the theoretical tools to
perform NLL resummation have been available for some
time~\cite{Kidonakis:1997gm} and a phenomenological study has been
performed in \cite{Kidonakis:2000gi}, it has recently been pointed out
that, for most jet definitions, jet cross sections are plagued by
nonglobal logarithms~\cite{Dasgupta:2003mk}, starting at NLL
level. Pushing the accuracy of soft-gluon resummation beyond leading
logarithms (LL) for these cross sections will thus require more work.

It seems fair to conclude that enough resummation technology exists to
perform a resummed global PDF fit. Including only DIS and Drell-Yan
data, such a fit could actually be consistently performed at NNLO/NNLL
level; in order to further constrain combinations of partons, which
are hard to determine using these data only, one might then decide to
trade some logarithmic accuracy in exchange for more data coverage.

In order to assess the impact that the inclusion of resummations might
have on parton distributions, and more specifically on large-$x$ quark
distributions, in the following we shall perform a fit of large-$x$
DIS data, using both NLO and NLL-resummed coefficient functions. It
should not be regarded as an attempt to a global fit (see, e.g.,
Ref.~\cite{Kuhlmann:1999sf} for an analysis of large-$x$ PDF's in the
context of a global fit), since we shall clearly be forced to make
several approximations in order to extract partons from such a
comparatively small data set. Rather, it should be seen as a toy model
of a resummed fit, providing a rough quantitative assessment of the
impact of resummations. We find that soft-gluon effects typically
suppress quark distributions by amounts ranging from a few percent to
about $15-20\%$ at large but not extreme values of $x$, $0.55 \leq x
\leq 0.75$, for moderate $Q^2$. Sum rules also force a compensating
enhancement in the distribution at smaller values of $x$, which,
however, cannot be reliably determined within our current
approximations. These effects would indeed warrant a more detailed
investigation, if the current goal for PDF-related uncertainties (a
few percent) were to be enforced also at these relatively large values
of $x$.

\sect{Data and parametrizations}
\label{dapar}

Large-$x$ DIS data come predominantly from fixed-target
experiments.  In order to have at our disposal different linear
combinations of large-$x$ partons, we shall consider here
charged-current (CC) data from neutrino-iron DIS, collected by the
NuTeV collaboration~\cite{Tzanov:2003gq,Naples:2003ne}, and
neutral-current (NC) data from muon scattering from the
NMC~\cite{Arneodo:1996qe} and
BCDMS~\cite{Benvenuti:1989rh,Benvenuti:1989fm} collaborations.

For our purposes, it will be sufficient to examine data at fixed
values of $Q^2$, which we shall pick not too small so as to minimize
the impact of power corrections, which are enhanced at the boundaries
of phase space. We also require good data coverage for all the
three experiments considered. We shall use $Q^2 = 31.62 ~{\rm GeV}^2$
and $Q^2 = 12.59 ~{\rm GeV}^2$, which corresponds to a cut in $W^2$
between $4$ and $5 ~{\rm GeV^2}$, given the measured values of $x$.
We shall check at the end that our results at the two selected values
of $Q^2$ are compatible with NLO perturbative evolution. Since
threshold resummation naturally takes place in Mellin moment space,
our procedure will be to construct parametrizations of the data at the
chosen values of $Q^2$, compute Mellin moments of the
parametrizations, and then use them to extract moments of the
corresponding PDF's, with and without resummation. The difference
between resummed and unresummed moments of PDF's is {\it per se} a
useful and solid result, since any QCD analysis can in principle be
reformulated in Mellin space.  In any case, we will also provide a
simple $x$-space parametrization in order to illustrate the impact of
the results in a more conventional manner. Studies of DIS structure
functions in moment space were also performed in \cite{simula}, by
making use of data from the CLAS detector at Jefferson Laboratory; the
corresponding values of $Q^2$ are however too small for a perturbative
study like the present one.

Let us now turn to the NMC, BCDMS and NuTeV data sets we are
considering.  An efficient and convenient parametrization of NMC and
BCDMS data for the NC structure function $F_2$, for proton, deuteron,
and separately for the nonsinglet combination, has been provided in
Ref.~\cite{Forte:2002fg}, and was recently upgraded for protons with
the inclusion of HERA data in Ref.~\cite{DelDebbio:2004qj}. The
parametrization was constructed by first generating a large set of
Monte Carlo copies of the original data, including all information on
errors and correlations; subsequently, a neural network was trained on
each copy of the data, yielding a set of parametrizations which, taken
together, give a faithful and unbiased representation of the
probability distribution in the space of structure functions.

In principle, the neural parametrization can be used for any values of
$x$ and $Q^2$. In practice, errors will become increasingly large when
one moves away from the region of the data. We use values of $Q^2$
which are well inside the measured region, with data coverage up to $x
= 0.75$.  Specifically, we will be interested in the nonsinglet
structure function $F_2^{\rm ns} (x, Q^2)$, which is unaffected by the
gluon contribution and provides a combination of quark distributions,
essentially $u - d$, which is linearly independent from the ones
sampled by NuTeV data. The neural parametrization of $F_2^{\rm ns} (x,
Q^2)$ was previously used~\cite{Forte:2002us} in conjunction with the
technique of truncated Mellin moments~\cite{Forte:1998nw} for a
determination of $\alpha_s$, which is unaffected by parametrization
biases.

\begin{figure}
\centerline{\resizebox{0.49\textwidth}{!}{\includegraphics{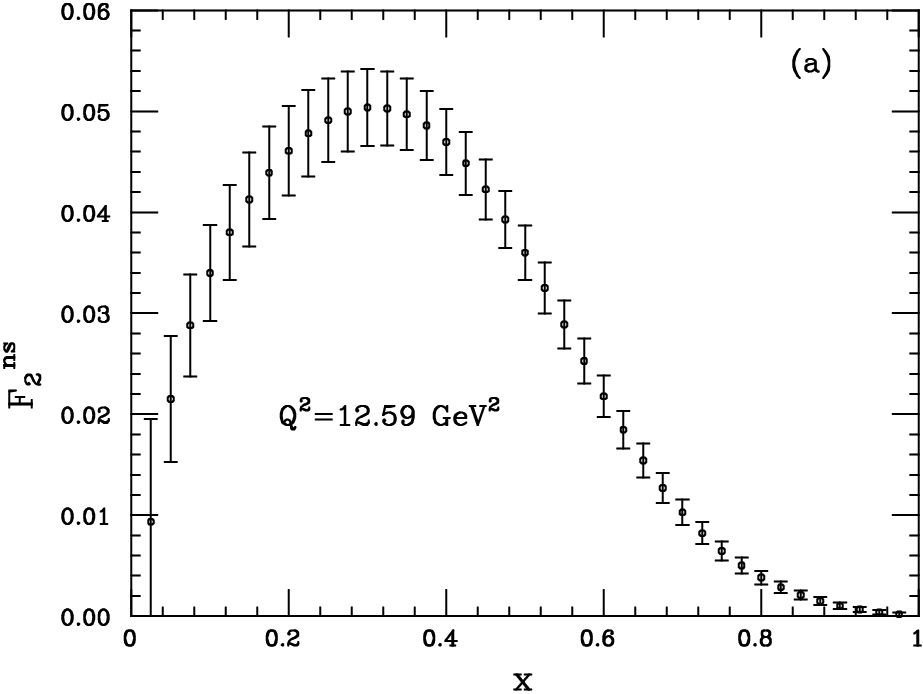}}%
\hfill%
\resizebox{0.49\textwidth}{!}{\includegraphics{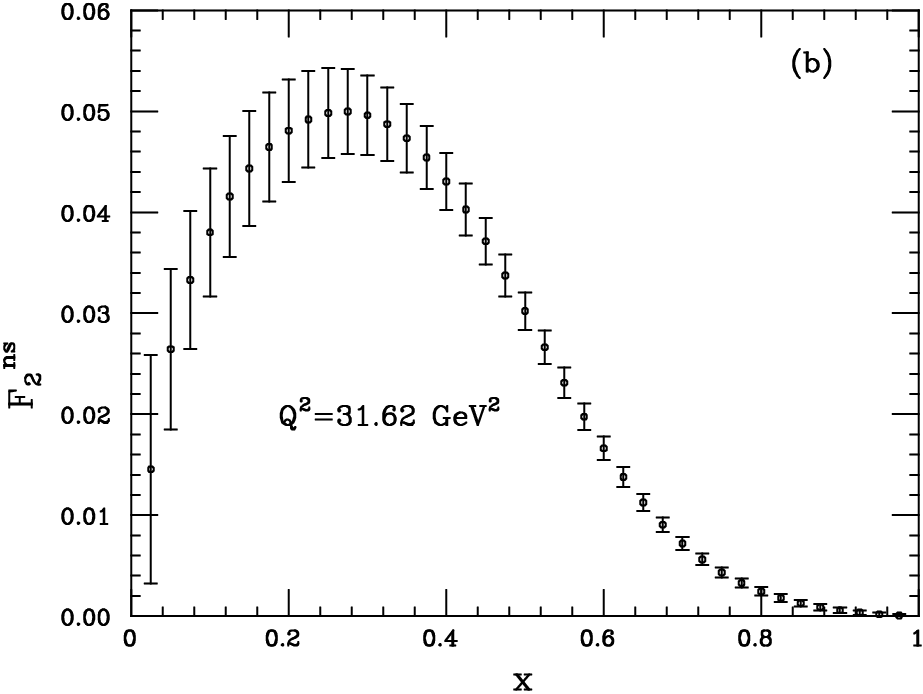}}}
\caption{A sampling of the neural parametrization of NMC and BCDMS
data for $F_2^{\rm ns} (x, Q^2)$ at $Q^2 = 12.59 ~{\rm GeV}^2$ (a) and
at $Q^2 = 31.62~{\rm GeV}^2$ (b), from the NNPDF
Collaboration~\cite{Forte:2002fg}.}
\label{figfor}
\end{figure}

To illustrate the quality of the data, we show in Fig.~\ref{figfor}
the nonsinglet structure function $F_2^{\rm ns} (x, Q^2)$, computed
with the neural parametrization at our chosen values of $Q^2$, and for
$x = n/40$, $n = 1, \ldots, 39$.  The central values are given by the
averages of the results obtained with the one thousand neural networks
of the NNPDF collaboration, and error bars are the corresponding
standard deviations. Error bars are relatively large, because
$F_2^{\rm ns} (x, Q^2)$ is the difference between proton and deuteron
structure functions, which entails a loss of precision. Central values
and errors for the moments are similarly obtained by computing the
moments with each neural network, and then taking averages and
standard deviations.

NuTeV provides data for the CC structure functions $F_2$ and
$F_3$. Since data are taken on an iron target, they need to be
rescaled to include nuclear corrections, which were computed in
\cite{seligman} by fitting the ratio $F_2^{Fe}/F_2^D$. The required
smearing factor is given by
\beq
N(x) = 1.10 - 0.36 \, x - 0.28 ~\exp( - 21.94 \, x ) + 2.77 \, x^{14.41}~.
\label{nucorr}
\eeq
We consider first the charged-current structure function $F_3$ and its
parton content. One has
\beq
x F_3 = \frac{1}{2} \left( x F_3^\nu + x F_3^{\bar \nu} \right) =
  x \left[\sum_{q,q'} |V_{qq'}|^2 \left( q - \bar q \right) \otimes 
  C_3^q \right]~,
\label{partF3}
\eeq
where $V_{qq'}$ are the relevant Cabibbo--Kobayashi--Maskawa (CKM)
matrix elements and $C_3^q$ is the appropriate coefficient function.
We fit the data at our chosen values of $Q^2$ using the functional
form
\beq
x F_3(x) = C x^{-\rho} ( 1 - x )^\sigma ( 1 + k x )~.
\label{fitF3}
\eeq
Eq.~(\ref{fitF3}) is quite similar to the functional form which is
used as initial condition for parton densities in the global analyses
\cite{Pumplin:2002vw, Martin:2002dr}. We checked the stability of our
fit by modifying the last factor of \Eq{fitF3} with the inclusion of
further powers of $x$ or logarithmic terms in $x$. We find that the
parametrization (\ref{fitF3}), with four tunable parameters, is reliable
enough to reproduce the data with quite small errors on the best-fit
parameters and reasonable values of the $\chi^2$ per degree of
freedom.

\begin{figure}
\centerline{\resizebox{0.49\textwidth}{!}{\includegraphics{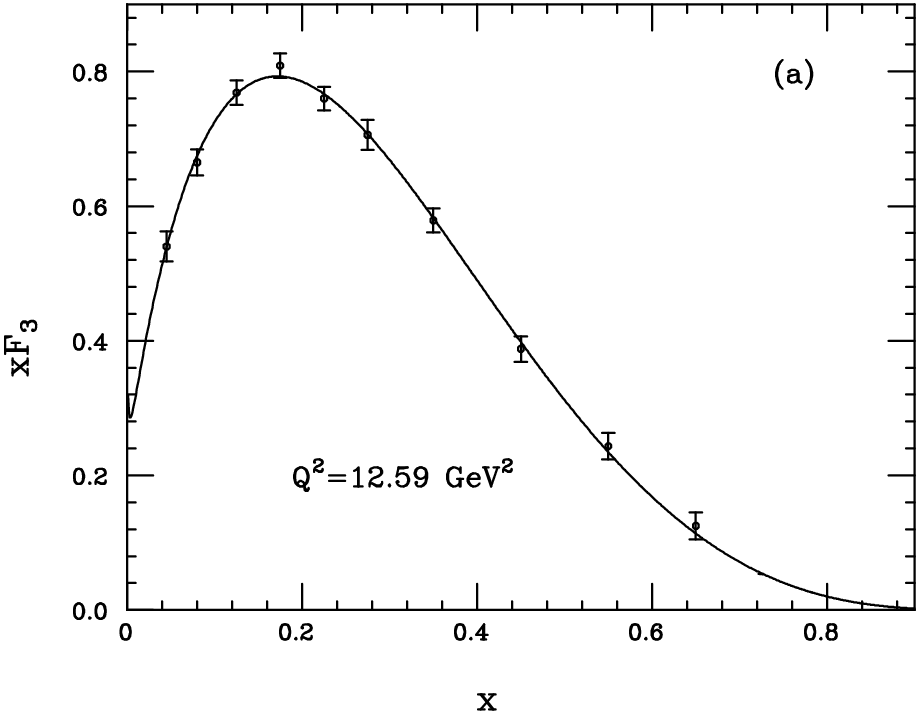}}%
\hfill%
\resizebox{0.49\textwidth}{!}{\includegraphics{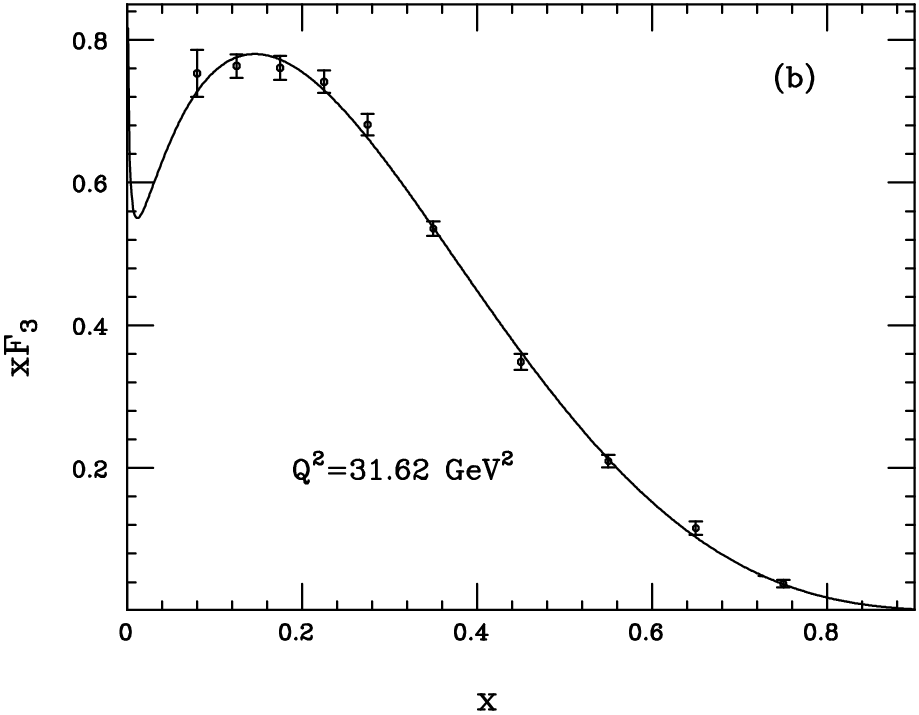}}}
\caption{NuTeV data on the structure function $x F_3$, at $Q^2 = 
12.59 ~{\rm GeV}^2$ (a) and at $Q^2 = 31.62 ~{\rm GeV}^2$ (b), 
along with the best
fit curve parametrized by \Eq{fitF3}.}
\label{figF3}
\end{figure}

The best-fit values at $Q^2 = 31.62$~GeV$^2$ are $C = 0.103 \pm
0.012$, $\rho = 0.294 \pm 0.034$, $\sigma = 3.325 \pm 0.089$, $ k =
42.972 \pm 4.700$, corresponding to $\chi^2/\mathrm{dof} = 7.20/6$. At
$Q^2 = 12.59$~GeV$^2$ we find instead $C = 0.054 \pm 0.005$, $\rho =
0.245 \pm 0.038$, $\sigma = 3.374 \pm 0.145$, $ k = 99.719 \pm 0.247$,
corresponding to $\chi^2/\mathrm{dof} = 2.06/6$.  The data and the
best-fit curves at the relevant values of $Q^2$ are shown in
Fig.~\ref{figF3}.

\begin{figure}
\centerline{\resizebox{0.49\textwidth}{!}{\includegraphics{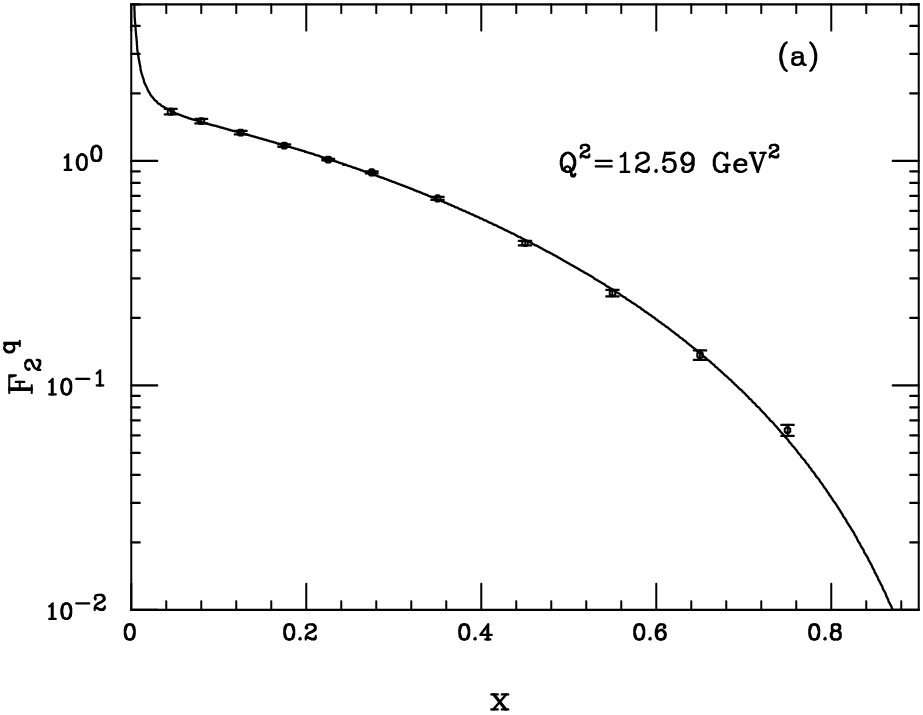}}%
\hfill%
\resizebox{0.49\textwidth}{!}{\includegraphics{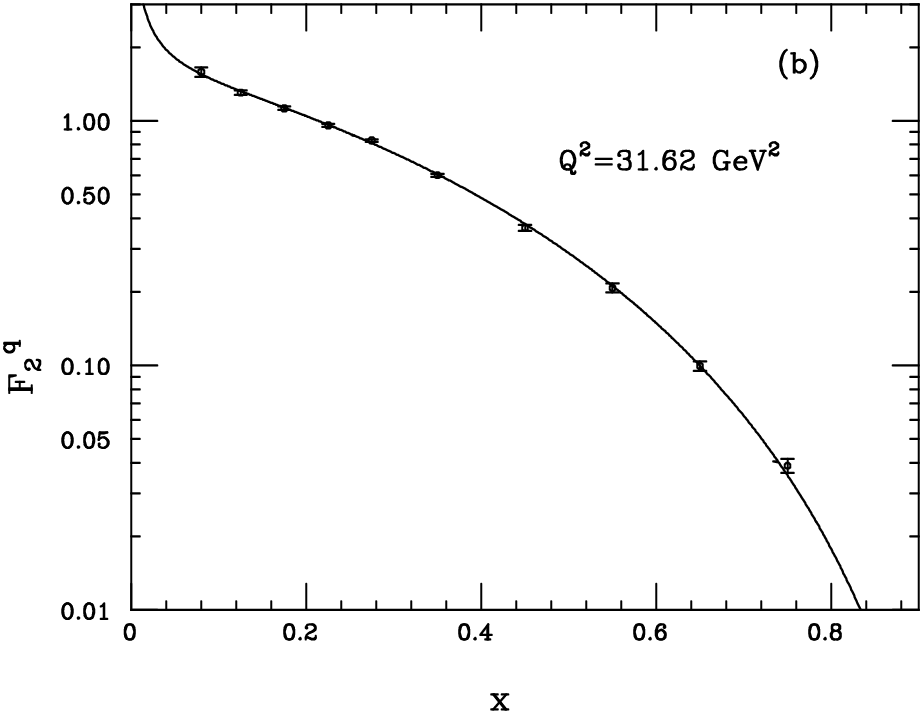}}}
\caption{NuTeV data on the quark-initiated contribution $F_2^q$ to the
structure function $F_2$, for $Q^2 = 12.59 ~{\rm GeV}^2$ (a), and $Q^2 =
31.62 ~{\rm GeV}^2$ (b). The solid lines are the best fit predictions
according to \Eq{fitF2}.}
\label{figF2}
\end{figure}

The situation for the structure function $F_2$, extracted from
charged-current data, is slightly more complicated since there is a
singlet component, and thus gluon-initiated processes also
contribute. Such processes are not logarithmically enhanced at
large $x$, and in fact, in the region of interest for our purposes,
the gluon contribution to the structure function is significantly
suppressed. We will handle it by subtracting it from the data point by
point, using a gluon distribution determined by a global fit. The
parton content of the charged-current structure function $F_2$ is
\beq
F_2 \equiv \frac{1}{2} \left( F_2^\nu + F_2^{\bar \nu} \right) = 
  x \sum_{q,q'} |V_{qq'}|^2 \left[(q + \bar q) \otimes C_2^q
  + g \otimes C_2^g \right] = F_2^q + F_2^g~.
\label{partF2}
\eeq
We will proceed by fitting only $F_2^q$ and computing the
gluon-initiated contribution using the gluon distribution from the NLO
set CTEQ6M \cite{Pumplin:2002vw}.  We have checked that our results
are not affected by the specific choice of gluon density, by repeating
the calculation with, e.g., the set MRST2001 \cite{Martin:2002dr}. As
above, we pick the parametrization
\beq
F_2^q (x) = F_2 (x) - F_2^g (x) = A \, x^{- \alpha} 
(1 - x)^\beta (1 + b x)~.
\label{fitF2}
\eeq
When doing the fit, we assume that we can neglect correlations among
data points, as well as the error on $F_2^g$ with respect to the error
on $F_2$ quoted by NuTeV. At $Q^2 = 31.62 ~{\rm GeV}^2$, the best fit
values for the parameters in Eq.~(\ref{fitF2}) are $A = 0.240 \pm
0.002$, $\alpha = 0.562 \pm 0.020$, $\beta = 3.211 \pm 0.065$, $b =
13.085 \pm 0.767$, with $\chi^2/\mathrm{dof}=9.99/6$. At $Q^2 = 12.59~
{\rm GeV}^2$, on the other hand, we find $A = 0.038 \pm 0.005$,
$\alpha = 0.816 \pm 0.021$, $\beta = 2.697 \pm 0.050$, $b = 66.804 \pm
7.583$, with $\chi^2/\mathrm{dof}=9.55/6$.  In Fig.~\ref{figF2} we
plot the data points of $F_2^q(x)$ at the chosen values of $Q^2$,
along with the curve given by \Eq{fitF2}, according to the central
values of the best-fit parameters.

At this point we have at our disposal parametrized expressions,
including errors and correlations for the parameters, within the
stated approximations, for the structure functions $x F_3$, $F_2^q$
and $F_2^{\rm ns}$. We can thus compute moments for the specified
values of $Q^2$, and extract the moments of the corresponding parton
densities by dividing out the appropriate coefficient functions, with
and without resummations.

\sect{A simple parton fit}
\label{sifit}

Having subtracted the contribution of gluon-initiated processes from
the charged current structure function $F_2$ in \Eq{partF2}, the
factorization
\beq
F_i (x, Q^2) = x \int_x^1 \frac{d \xi}{\xi} \, q_i \left(\xi, \mu_F^2 
  \right) C_i \left( \frac{x}{\xi}, \frac{Q^2}{\mu_F^2},
  \alpha_s (\mu_R^2) \right)
\label{myfact}
\eeq
applies to all the structure functions we shall be considering ($F_i =
\{F_2, xF_3, F_2^{\rm ns}\}$). In all cases $q_i$ is a combination of
(anti)quark distributions, while $C_i$ is the appropriate coefficient
function.

The coefficient functions $C_i$ for quark-initiated DIS contain terms
that become large when the Bjorken variable $x$ for the partonic
process is close to $x = 1$, which forces gluon radiation from the
incoming quark to be soft or collinear.  At ${\cal O}(\alpha_s)$, for
example, the coefficient functions can be written in the form,
\beq 
 C_i^{\rm NLO} \left(x,\frac{Q^2}{\mu_F^2},\alpha_s(\mu_R^2) \right) =
 \delta(1 - x) + {{\alpha_s(\mu_R^2)}\over{2\pi}} H_i
 \left(x,\frac{Q^2}{\mu_F^2} \right) \, .
\label{nlocoeff}
\eeq
Treating all quarks as massless, the part of $H_i$ which
contains terms that are logarithmically enhanced as $x \to 1$ reads:
\beq
 H_{i, {\mathrm{soft}}} \left(x,\frac{Q^2}{\mu_F^2} \right)
 = 2 C_F \left\{ \left[ {{\ln(1 - x)}\over{1 - x}} \right]_+ +
 {1 \over{(1 - x)_+}} \left( \ln{{Q^2} \over{\mu_F^2}} - 
 {3 \over 4} \right) \right\} \, .
\label{nlolargex}
\eeq
Taking a Mellin transform, the contributions proportional to $\alpha_s
[\ln(1 - x)/(1 - x)]_+$ and to $\alpha_s [1/(1 - x)]_+$ correspond to
double $(\alpha_s \ln^2 N)$ and single $(\alpha_s \ln N)$ logarithms
of the Mellin variable $N$.  Retaining only terms that are singular at
large $N$ one finds in fact
\beq
 \hat{H}_{i, {\mathrm{soft}}} \left(N, \frac{Q^2}{\mu_F^2} 
 \right) = 2 C_F \left\{ {1\over 2} \ln^2 N + \left[ \gamma_E + 
 {3 \over 4} - \ln{{Q^2}\over{\mu_F^2}} \right] \ln N \right\} \, .
\label{nlolargen}
\eeq

The resummation of soft-gluon effects, responsible for this singular
behaviour of DIS structure functions, has been well understood for a
long time \cite{Sterman:1986aj,Catani:1989ne}: it results in
exponentiation of all singular contributions to the Mellin moments of
the coefficient functions $C_i$ at large values of the moment variable
$N$. In the $\overline{\mathrm{MS}}$ factorization scheme, soft
resummation was implemented in \cite{Catani:1990rr,Vogt:2000ci} in the
massless approximation, and in \cite{Corcella:2003ib,Laenen:1998kp}
for heavy quark production.  In the following, we shall consider
values of $Q^2$ much larger than the relevant quark masses, so that we
can safely apply the results in the massless approximation.

The pattern of exponentiation of logarithmic singularities is
nontrivial: one finds that Mellin moments of the coefficient functions
can be written as
\beq
 \hat{C}_i^{\rm res} \left(N,\frac{Q^2}{\mu_F^2},\alpha_s(\mu_R^2) \right) =
 {\cal R} \left(N,\frac{Q^2}{\mu_F^2},\alpha_s(\mu_R^2) \right) 
 \Delta \left(N,\frac{Q^2}{\mu_F^2},\alpha_s(\mu_R^2) \right) \, ,
\label{genexpon}
\eeq
where ${\cal R}$ is a finite remainder, nonsingular as $N \to \infty$,
while~\cite{Sterman:1986aj,Catani:1989ne,Forte:2002ni}
\beqa
 \ln \Delta \left(N,\frac{Q^2}{\mu_F^2},\alpha_s(\mu_R^2) \right) & = & 
 \int_0^1 {dx \, {{x^{N - 1} - 1}\over{1 - x}}}
 \Bigg\{ \int_{\mu_F^2}^{(1 - x) Q^2} {{d k^2} \over {k^2}}
 A \left[ \alpha_s(k^2) \right] \nonumber \\ 
 & + & B \left[ \alpha_s \left(Q^2 (1 - x) 
 \right) \right] \Bigg\} \, .
\label{disdelta}
\eeqa
In \Eq{disdelta} the leading logarithms (LL), of the form $\alpha^n_s
\ln^{n + 1} N$, are generated at each order by the function $A$.
Next-to-leading logarithms (NLL), on the other hand, of the form
$\alpha_s^n \ln^n N$, require the knowledge of the function $B$.  In
general, resumming ${\rm N}^k{\rm LL}$ to all orders requires the
knowledge of the function $A$ to $k+1$ loops, and of the function $B$
to $k$ loops. In the following, we will adopt the common standard of
NLL resummation, therefore we need the expansions
\beq
 A(\alpha_s) = \sum_{n = 1}^{\infty} \left({{\alpha_s} \over
 {\pi}} \right)^n A^{(n)}\ \ ;\ \ 
 B(\alpha_s) = \sum_{n = 1}^{\infty} \left({{\alpha_s} \over
 {\pi}} \right)^n B^{(n)}
\label{nllexpan}
\eeq
to second order for $A$ and to first order for $B$. The relevant
coefficients are
\beqa
 A^{(1)} & = & C_F \, , \nonumber \\
 A^{(2)} & = & {1 \over 2} C_F \left[ C_A \left(
 {{67} \over{18}} - {{\pi^2} \over 6} \right) - {5 \over 9} n_f 
 \right] \, , \\
 B^{(1)} & = & - {3 \over 4} C_F \, . \nonumber
\label{nllresco}
\eeqa
Notice that in \Eq{disdelta} the term containing the function
$A(\alpha_s)$ resums the contributions of gluons that are both soft
and collinear, and in fact the anomalous dimension $A$ can be
extracted order by order from the residue of the singularity of the
nonsinglet splitting function as $x \to 1$.  The function $B$, on the
other hand, is related to collinear emission from the final state
current jet.  In the case of heavy quarks, the function $B(\alpha_s)$
needs to be replaced by a different function, called $S(\alpha_s)$ in
\cite{Corcella:2003ib}, which is instead characteristic of processes
with massive quarks, and includes effects of large-angle soft
radiation.

Turning to our fit, we observe that, upon taking Mellin moments, the
convolution in \Eq{myfact} turns into a product, and it becomes
straightforward to extract moments of the parton combinations $q_i (x,
Q^2)$ at NLO, or with NLL resummation. Setting $\mu_F = \mu_R = Q$,
one simply finds
\beq
 \hat{q}_i^{\mathrm{NLO}}(N, Q^2) = \frac{\hat{F}_i(N - 1, 
 Q^2)}{\hat{C}_i^{\mathrm{NLO}} \left(N, 1, \alpha_s(Q^2) \right)}
 \, \, ; \, \, \, \, 
 \hat{q}_i^{\mathrm{res}}(N, Q^2) = \frac{\hat{F}_i(N - 1, 
 Q^2)}{\hat{C}_i^{\mathrm{res}} \left(N, 1, \alpha_s(Q^2) \right)}~,
\label{impres}
\eeq
where the resummed coefficient function has been suitably matched to
NLO, in order to avoid double counting of logarithmic contributions.

Since we are considering only three measurements, we need to introduce
further approximations in order to be able to extract individual
parton distributions. We will use isospin symmetry of the sea, so that
$\bar{u} = \bar{d}$ and $s = \bar{s}$; further, we shall take the
charm quark distribution to vanish and, for simplicity, we will impose
a simple proportionality relation between antiquark distributions,
$\bar{s} = \kappa \, \bar{u}$. In the fit shown below, we shall assume
$\kappa =1/2$.  All of these assumptions are essentially harmless at
large $x$, where sea quarks are negligible: they allow us, however, to
solve for the valence quark distributions $u$, $d$ and, say, $s$. The
expressions for the parton combinations $q_i$ become particularly
simple if one approximates the elements of the CKM matrix by
neglecting terms of order $(\sin \theta_C)^4$ in $|V_{qq'}|^2$. Within
the stated approximations, one finds then
\beqa
q_2 (x, Q^2) & = & u (x, Q^2) + d (x, Q^2) + 3 \, s (x, Q^2)~, \nonumber \\
q_3 (x, Q^2) & = & u (x, Q^2) + d (x, Q^2) - s (x, Q^2)~ \label{qdist} \\
q_2^{\mathrm{ns}} (x, Q^2) & = & \frac{1}{6} \left(u (x, Q^2) - d (x, Q^2) 
  \right)~, \nonumber
\eeqa
which is easily inverted to give $u$, $d$ and $s$.
\begin{figure}
\centerline{\resizebox{0.65\textwidth}{!}{\includegraphics{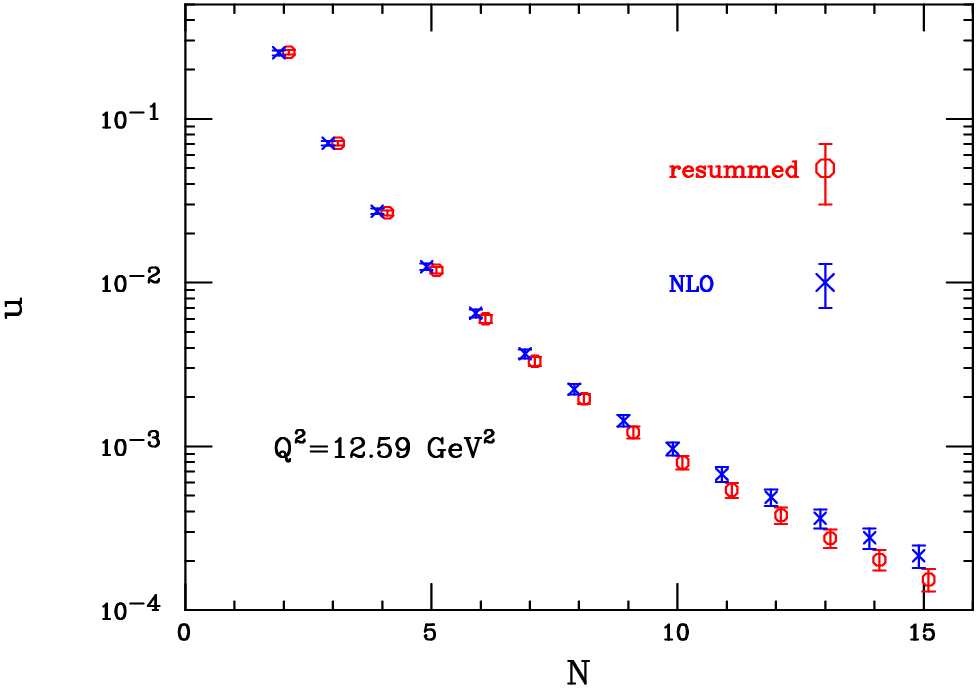}}}
\caption{Moments of the resummed and NLO up quark distribution with
statistical errors, in the range $2 \leq N \leq 15$, at 
$Q^2=12.59$~GeV$^2$.}
\label{figun1}
\end{figure}
\begin{figure}
\centerline{\resizebox{0.65\textwidth}{!}{\includegraphics{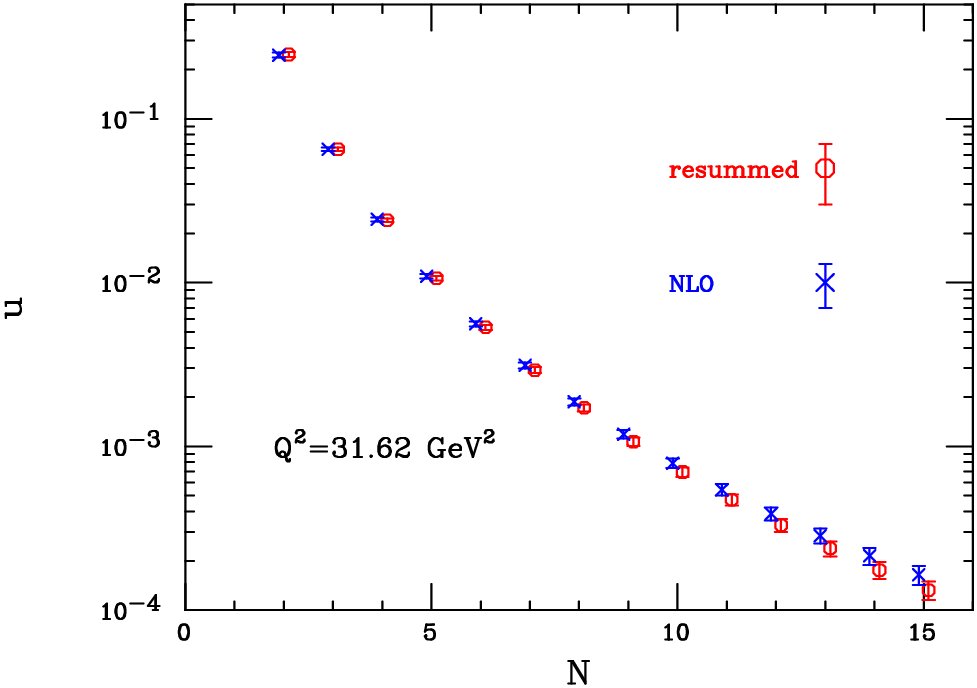}}}
\caption{As in Fig.~\ref{figun1}, but at $Q^2=31.62$~GeV$^2$.}
\label{figun2}
\end{figure}
Having extracted the moments of the parton combinations in \Eq{qdist},
one easily derives moments of individual quark distributions. In order
to estimate the error on the moments, we proceed as follows.
Regarding $F_2^{\rm ns}$, the neural parametrization is designed to
allow for a simple calculation of the error on any functional of the
data: one simply computes the standard deviation of the desired
observable over the set of neural nets. A similar procedure would also
yield correlations between different observables (in our case
moments). In the present case, we shall neglect correlations between
moments since we do not have a sufficiently reliable method to
evaluate them in the case of NuTeV data.  Concerning charged-current
structure functions, we mimick the neural method by generating a Monte
Carlo set of parametrizations of $F_2$ and $F_3$ at the chosen values
of $Q^2$, assuming the parameters of our fits are gaussian distributed
around their mean values with the stated errors. We then compute
errors on moments as standard deviations over the Monte Carlo set.

Our results for the moments of the valence up quark distribution are
shown in Figs.~\ref{figun1}--\ref{figun2}.  The trend is clear, as
could have been anticipated by comparing the resummed coefficient
function with the NLO one: in the $\overline{\mathrm{MS}}$ scheme,
resummation enhances the moments of the coefficient function, and thus
suppresses the moments of quark distributions, with an effect
increasing with the moment index $N$. The effect is unequivocal at
both values of $Q^2$, since resummed and NLO moments differ by more
than their error, beginning with $N \sim 6$. At $Q^2 = 31.62$~GeV$^2$ errors
are somewhat smaller, however the effect of resummation is also
slightly reduced, so that errors tend to have a somewhat larger
overlap. In fact, as observed, e.g., in Ref.~\cite{Corcella:2003ib},
LL and NLL terms in the Sudakov exponent are weighted by powers of
$\alpha_s$ and the coupling constant is larger when evaluated at a
lower scale.  The trend is, in any case, still clearly visible, and
the shift in the central values is of comparable size.

The down quark distribution is significantly smaller than the up quark
distribution at large $x$, thus large-$N$ moments are suppressed. With
our current, relatively small data set, the effect of resummation on
the down quark distribution is completely overshadowed by statistical
errors, so that the values of the moments are compatible with $0$
beginning at $N \sim 10$. Similarly, moments of the strange quark
distribution cannot be reliably determined with these data.  We shall
then concentrate on the up quark distribution for the remainder of our
analysis. One may just note in passing that resummation does not
appear to shift even the central values of the moments of $d(x, Q^2)$:
taken seriously, this would suggest that the bulk of the effect is
carried by the largest valence distribution for the chosen hadron.

\begin{figure}[ht!]
\centerline{\resizebox{0.65\textwidth}{!}{\includegraphics{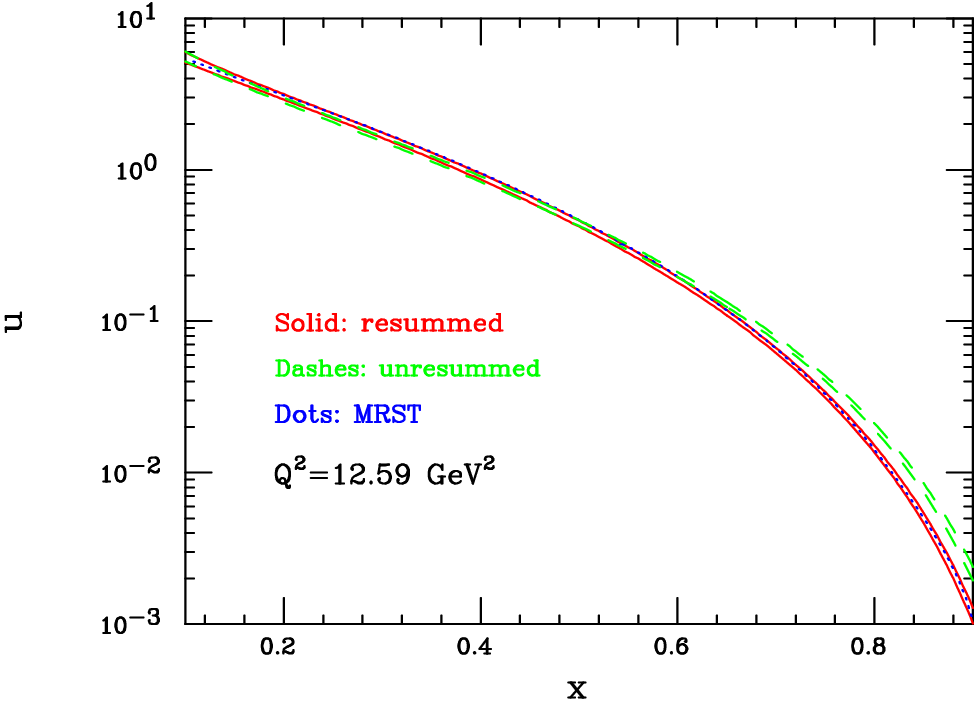}}}
\caption{Resummed and NLO up quark distributions, at
$Q^2=12.59$~GeV$^2$, reconstructed by fitting moments with a simple
functional form, $u(x) = D x^{-\gamma} (1 - x)^\delta$. The MRST2001
NLO up quark distribution at the appropriate $Q^2$ is included for
comparison. The plot shows the edges of a band at the
one-standard-deviation confidence level.}
\label{figux1}
\end{figure}

\begin{figure}[ht!]
\centerline{\resizebox{0.65\textwidth}{!}{\includegraphics{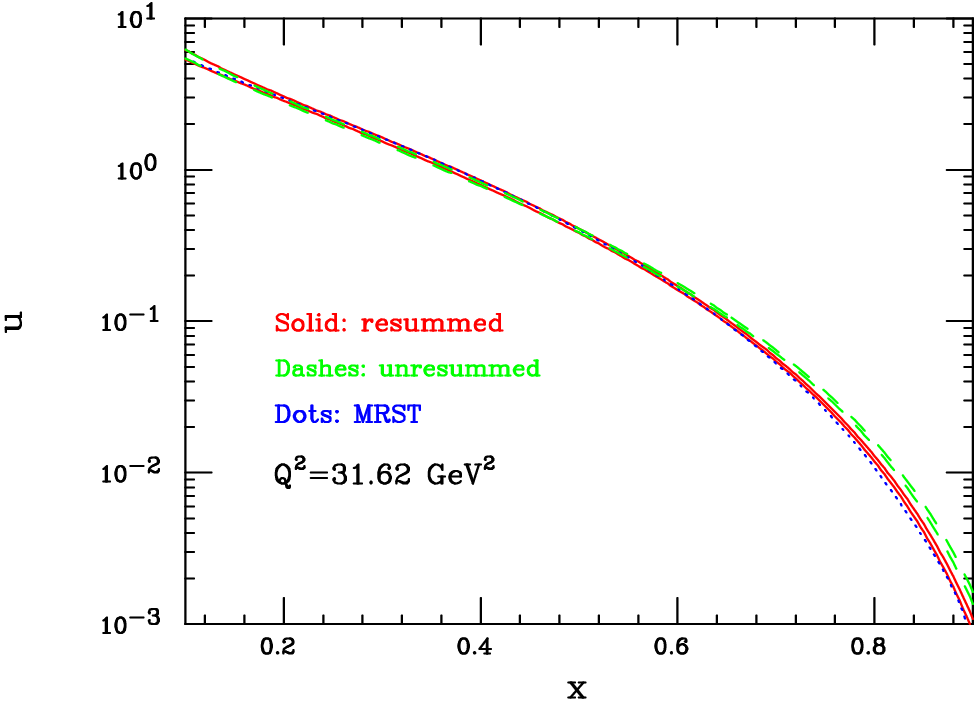}}}
\caption{As in Fig.~\ref{figux1}, but at $Q^2=31.62$~GeV$^2$.}
\label{figux2}
\end{figure}

In order to provide a more conventional (and possibly a more
practical) parametrization of the effect of resummations, we have also
performed a fit of the moments presented in
Figs.~\ref{figun1}--\ref{figun2} to a simple $x$-space
parametrization, choosing:
\begin{equation}
 u (x) = D \, x^{-\gamma} \left(1 - x \right)^\delta \, .
\label{uxspace}
\end{equation}  
The functional form (\ref{uxspace}) is sufficient to fit the moments
of the up quark distribution with small errors on $D$, $\gamma$ and
$\delta$, and a low $\chi^2/\mathrm{dof}$. We have checked that the
inclusion of other terms, for example a further factor linear in $x$,
as in Eqs.~(\ref{fitF3}) and (\ref{fitF2}) does not significantly
improve the quality of the fit. The results using the fitting function
(\ref{uxspace}) are shown in Figs.~\ref{figux1}--\ref{figux2}.  The
best-fit parameters at $Q^2 = 12.59$ GeV$^2$ are $D = 3.025 \pm
0.534$, $\gamma = 0.418\pm 0.101$ and $\delta = 3.162 \pm 0.116$, with
$\chi^2/\mathrm{dof}=1.62/11$, for the NLO fit, while the resummed fit
yields $D = 4.647 \pm 0.881$, $\gamma = 0.247 \pm 0.109$ and $\delta =
3.614 \pm 0.128$, with $\chi^2/\mathrm{dof}=0.64/11$.  At $Q^2 =
31.62$ GeV$^2$ we find $D = 2.865 \pm 0.420$, $\gamma = 0.463 \pm
0.086$, $\delta = 3.301 \pm 0.098$ and $\chi^2/\mathrm{dof}=1.10/11$
for the NLO fit, as well as $D = 3.794 \pm 0.583$, $\gamma = 0.351 \pm
0.090$, $\delta = 3.598 \pm 0.104$ and $\chi^2/\mathrm{dof}=0.53/11$
for the resummed fit\footnote{We note that the values of
$\chi^2/\mathrm{dof}$ are very small. This might be due to the fact
that we are neglecting correlations between moments.}.  The error
bands in Figs.~\ref{figux1}--\ref{figux2} correspond to a prediction
at one-standard-deviation confidence level.  They are obtained, as
above, by generating a Monte Carlo sample of parametrizations of the
stated form, assuming a gaussian distribution for the parameters $D$,
$\gamma$ and $\delta$ with the stated errors; thus, they reflect only
statistical errors and do not take into account biases due to the
simple choice of functional form.

\begin{figure}
\centerline{\resizebox{0.49\textwidth}{!}{\includegraphics{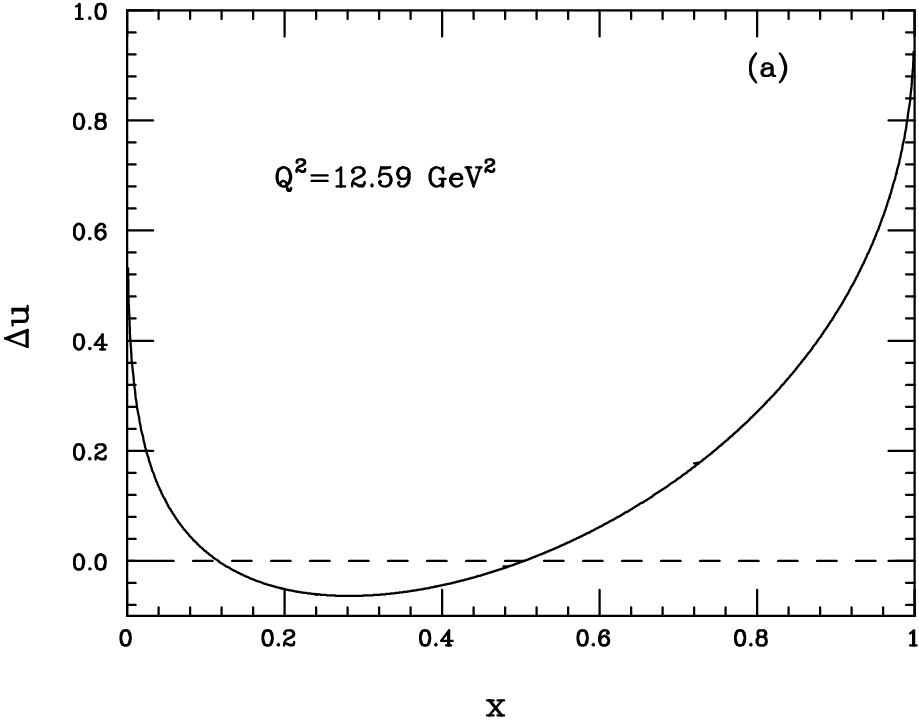}}%
\hfill%
\resizebox{0.49\textwidth}{!}{\includegraphics{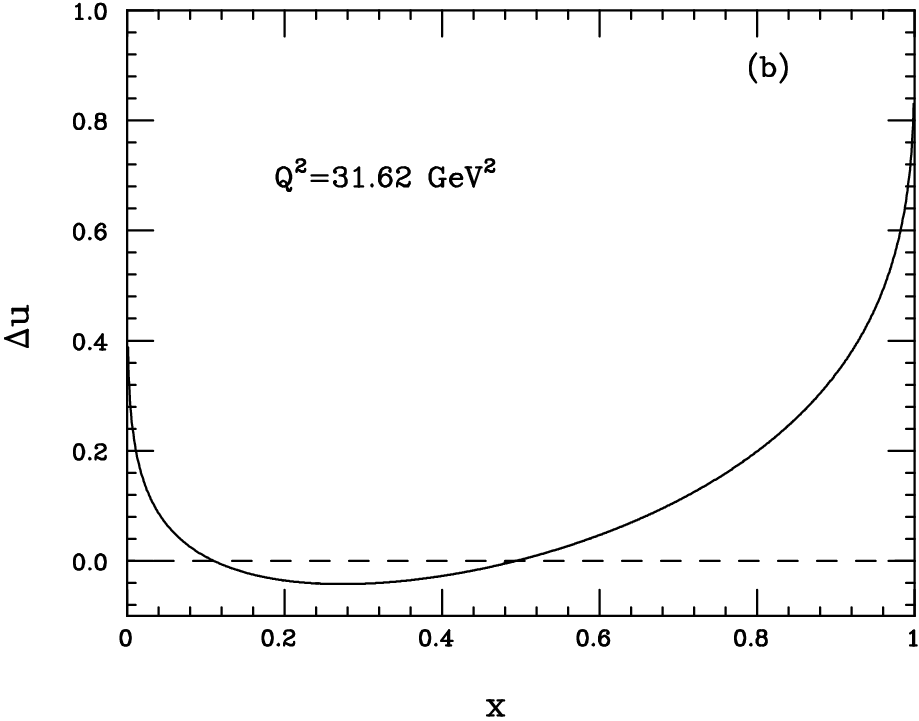}}}
\caption{Central value for the relative change in the up quark distribution, 
$\Delta u (x) \equiv \left(u_{\rm NLO} (x) - u_{\rm res} (x) 
\right)/u_{\rm NLO} (x)$, at $Q^2=12.59$ (a) and 31.62~GeV$^2$ (b).}
\label{figudel}
\end{figure}

To display more clearly the quantitative effect of the resummation, we
also present in Fig.~\ref{figudel} the central values for the
normalized deviation of the resummed prediction from the NLO
distribution, $\Delta u(x) = \left(u_{\rm NLO} (x) - u_{\rm res} (x)
\right)/u_{\rm NLO} (x)$, at the two chosen values of $Q^2$.


A few comments concerning Figs.~\ref{figux1}--\ref{figudel} are in
order. First of all, an evident feature of the result is the change in
sign of the effect around the point $x = 1/2$. This is a stable
feature of all our fits, and it can be traced back to the momentum sum
rule, which is essentially unaffected by the resummation. Depletion of
valence quarks at large $x$ is thus partly compensated by an increase
at smaller values of $x$. The further change in sign at values of $x$
around 0.1, on the other hand, cannot be taken too seriously, since it
happens in a region which is dominated by extrapolation within our
current data set, so that errors are correspondingly very large. The
impact of resummation is larger, as must be expected, at the lower
value of $Q^2$.

At $Q^2 = 12.59$ GeV$^2$ and large $x$, it is to be expected that
power corrections will play a role too. We have chosen, however, not
to introduce them explicitly in our parametrization of $F_i (x, Q^2)$
since, as discussed in the introduction, their effect is inevitably
tied to the precise treatment of the resummation. Disentangling
resummations and power corrections is best left to a more precise
quantitative analysis performed in the context of a global fit. We
have, in any case, checked that target mass corrections do not
significantly influence our results.

\begin{figure}[ht!]
\centerline{\resizebox{0.65\textwidth}{!}{\includegraphics{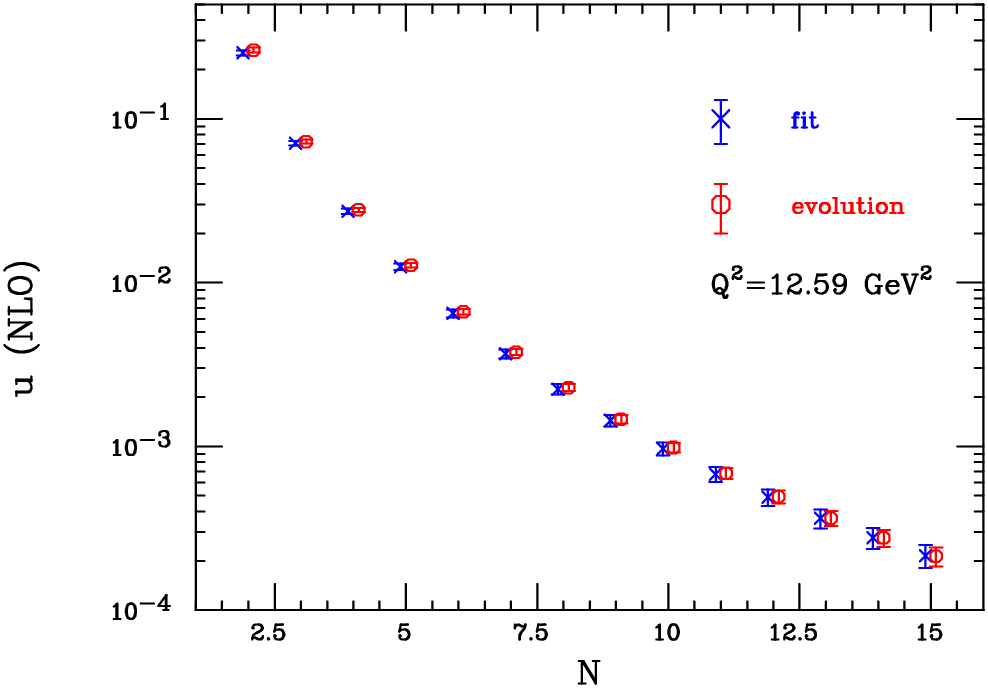}}}
\caption{Comparison of fitted moments of the NLO 
up quark distribution, at $Q^2 =
12.59$~GeV$^2$, with moments obtained via NLO evolution from 
$Q^2 = 31.62$~GeV$^2$.}
\label{figev1}
\end{figure}

\begin{figure}[ht!]
\centerline{\resizebox{0.65\textwidth}{!}{\includegraphics{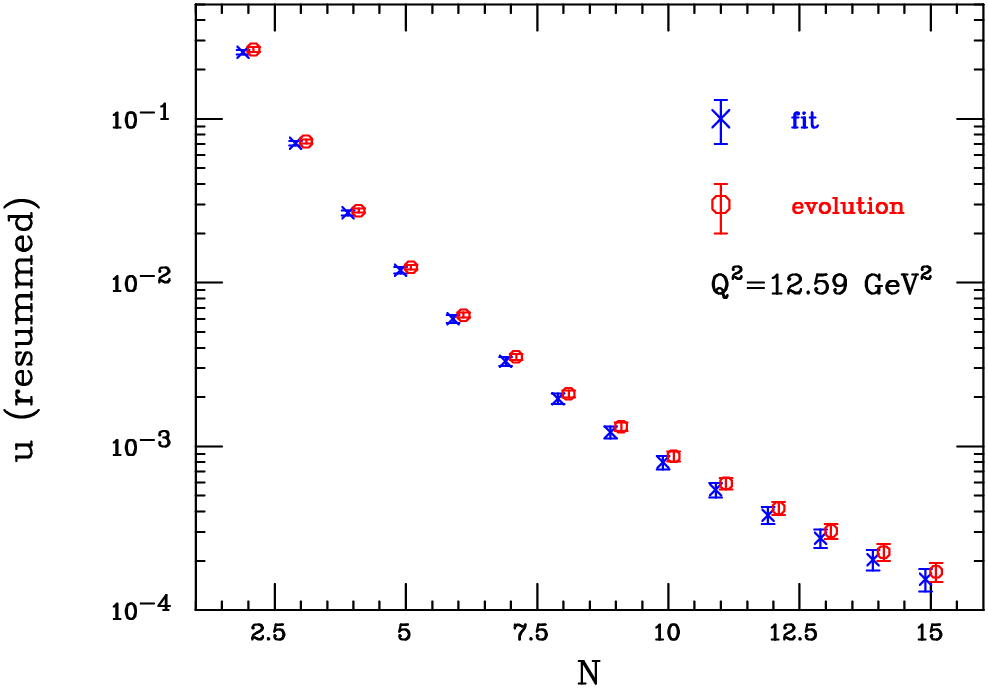}}}
\caption{As in Fig.~\ref{figev1}, but comparing NLL-resummed moments
of the up quark density.}
\label{figev2}
\end{figure}

We have verified that our fits, which are performed independently at
two fixed values of $Q^2$, are consistent with perturbative
evolution. To this end, we have evolved the moments of the up quark
distribution, starting at $Q^2 = 31.62$~GeV$^2$, using the NLO
Altarelli--Parisi anomalous dimension, down to $Q^2 = 12.59$~GeV$^2$,
and we have compared the results of the evolution with the direct fits
of Figs.~\ref{figun1}--\ref{figun2}.  The comparison is shown in
Figs.~\ref{figev1}--\ref{figev2}.  The results are fully compatible
within errors, and we believe that the agreement could be further
improved if one included power corrections, which are relevant
especially at $Q^2=12.59$~GeV$^2$.  In the evolution, we have used
$\alpha_s (12.59\ {\rm GeV}^2) = 0.2394$, corresponding to $\alpha_s
(31.62\ \rm{GeV}^2) = 0.2064$, or to $\Lambda_{\rm QCD}^{(5)} = 226$
MeV , with appropriate matching at the $b$-quark mass threshold, set
to $m_b =$ 4.5 GeV.  Our choices are consistent with
Ref.~\cite{Pumplin:2002vw}.

From a phenomenological point of view, we see that the impact of
soft-gluon resummation on quark distributions can be sizeable, albeit
only at values of $x$ which are quite large, say $0.55 < x < 0.75$.
More precisely, as one can verify from Fig.~\ref{figudel}, the
suppression of the resummed up quark distribution with respect to the
NLO one reaches $5\%$ at $x \simeq 0.58$, $10\%$ at $x \simeq 0.65$
and $20\%$ at $x \simeq 0.75$ for $Q^2 = 12.59$ GeV$^2$, while for
$Q^2 = 31.62$ GeV$^2$ the same suppression factors are reached at $x
\simeq 0.61$, $x \simeq 0.69$ and $x \simeq 0.8$, respectively.  Such
values of $x$ can be relevant for several high energy processes,
ranging from the production of high-mass Drell--Yan pairs, to
high-$E_T$ jets, to the exchange of heavy resonances in the
$t$-channel of hadron collisions. Considering, for example, the eccess
of high-$E_T$ jets seen at the Tevatron Run I by
CDF~\cite{Affolder:2001fa}, the effect of including resummations in a
PDF fit would have been to actually enhance the discrepancy between
theory and experiment, since resummations suppress valence quarks at
large $x$, and thus would have lowered the QCD prediction. Of course,
a fully consistent treatment would have required making use of a
resummed partonic cross section as well, which might have had a
balancing effect. Interestingly, resummation may well be moving
valence quarks from large to medium values of $x$, though the evidence
for that in our present fit is at best qualitative. If that were the
case, one might encounter several competing effects, depending on the
partonic subprocess. For example, in Drell-Yan production at high mass
and high rapidity, the heavy vector boson is produced by fusion of
high-$x$ and low-$x$ partons, and one could have a depletion when the
high-$x$ parton is a quark, or a slight enhancement (or no effect at
all) when the high-$x$ parton is a gluon and the quark has lower
momentum fraction. Finally, it should be noted that, in a fully
consistent treatment, the depletion of the cross section due to
resummed PDF's may well compete against the enhancement of the hard
partonic cross section which is commonly found when resumming
logarithms to that accuracy, resulting in a reduced impact of soft
gluons at hadron level.

\sect{Outlook}
\label{outlo}

We have performed a first attempt to assess the impact of soft-gluon
resummation on fits of parton distribution functions.  We have argued
that a global fit including soft-gluon effects is both feasible and
desirable, from a theoretical as well as phenomenological point of
view. That being said, it is useful to gauge the size of the effect
that resummation might have.  To that end, we have performed a simple
fit of large-$x$ DIS data from the NuTeV, NMC and BCDMS
collaborations. Our fit is not meant to be used as a complement or a
substitute for a global fit: it is based on a small set of data,
concentrated at large $x$, and does not consistently include all the
constraints arising from sum rules and evolution which are properly
taken into account in global fits. Our results should instead be seen
as a first semiquantitative study of the impact of soft-gluon effects,
and we believe that they might be an incentive for the dedicated
collaborations performing PDF fits to include these effects in their
algorithms. We have shown that, in the $\overline{\mathrm{MS}}$
scheme, the main effect of soft resummation is to suppress valence
quark distributions at large $x$, by an amount ranging from a few
percent to as much as 20$\%$, in the range $0.55 < x < 0.75$.  This
suppression may be partly compensated by a weaker enhancement, of the
order of a few percent, at medium values of $x$, $0.1 < x < 0.5$, an
effect which however falls largely inside our current statistical
errors. It should be noted that a sizeable effect of this kind at
large values of $x$ and moderate $Q^2$ will feed through to smaller
values of $x$ at large $Q^2$ via evolution. We expect that including
resummations should help to lower the theoretical uncertainty in the
determination of PDF's, and more in general in QCD cross sections,
both by reducing scale uncertainties, and by allowing for stronger
constraints on large-$x$ partons, thanks to the inclusion of more data
points. We note also that the effect of resummations on PDF's (a
suppression) may turn out to be competing with the effect on partonic
cross sections (in general, an enhancement). Disentangling such
competing effects to gain a precise quantitative understanding of soft
gluon effects on hadron-level cross sections can only be achieved with
a consistent treatment of resummations, including their effects both
in PDF global fits and in hard cross sections.

\vspace{1cm}

{\large \bf Acknowledgements}

\noindent The authors would like to thank S. Forte, M.L. Mangano and
M.H. Seymour for useful discussions, and L. Del Debbio, J. Rojo and
A. Piccione for help with the neural network code of the NNPDF
collaboration.  We are also grateful to the NuTeV collaboration, and
especially to D. Naples and M. Tzanov, for their assistance in the use
of the NuTeV structure function data.  L.M. wishes to thank CERN for
hospitality and support during the early phase of this work, which is
also supported in part by MIUR under contract $2004021808\_009$.


\begin{thebibliography}{99}

\bibitem{Giele:2002hx}
  W.~Giele {\it et al.}, in {\it Les Houches 2001}, proceedings of the
  Workshop ``Physics at TeV colliders'', 275, {\tt hep-ph/0204316}.

\bibitem{Pumplin:2002vw}
  J.~Pumplin, D.~R.~Stump, J.~Huston, H.~L.~Lai, P.~Nadolsky and W.~K.~Tung,
  {\it JHEP} {\bf 0207} (2002) 012,
  {\tt hep-ph/0201195}.

\bibitem{Martin:2002dr}
  A.~D.~Martin, R.~G.~Roberts, W.~J.~Stirling and R.~S.~Thorne,
  {\it Phys. Lett.} {\bf B 531} (2002) 216,
  {\tt hep-ph/0201127}.

\bibitem{Alekhin:2002fv}
  S.~Alekhin,
  {\it Phys. Rev.} {\bf D 68} (2003) 014002,
  {\tt hep-ph/0211096}.

\bibitem{Vogt:2004mw}
  A.~Vogt, S.~Moch and J.~A.~M.~Vermaseren,
  {\it Nucl. Phys.} {\bf B 691} (2004) 129,
  {\tt hep-ph/0404111}.

\bibitem{Moch:2004pa}
  S.~Moch, J.~A.~M.~Vermaseren and A.~Vogt,
  {\it Nucl. Phys.} {\bf B 688} (2004) 101,
  {\tt hep-ph/0403192}.

\bibitem{Sterman:1986aj}
  G.~Sterman,
  {\it Nucl. Phys.} {\bf B 281} (1987) 310.

\bibitem{Catani:1989ne}
  S.~Catani and L.~Trentadue,
  {\it Nucl. Phys.} {\bf B 327} (1989) 323.

\bibitem{Banfi:2004yd}
  A.~Banfi, G.~P.~Salam and G.~Zanderighi,
  {\it JHEP} {\bf 0503} (2005) 073,
  {\tt hep-ph/0407286}.

\bibitem{Martin:1998np}
  A.~D.~Martin, R.~G.~Roberts, W.~J.~Stirling and R.~S.~Thorne,
  {\it Phys. Lett.} {\bf B 443} (1998) 301,
  {\tt hep-ph/9808371}.

\bibitem{Gardi:2002xm}
  E.~Gardi and R.~G.~Roberts,
  {\it Nucl. Phys.} {\bf B 653} (2003) 227,
  {\tt hep-ph/0210429}.

\bibitem{Sterman:2000pu}
  G.~Sterman and W.~Vogelsang,
  {\tt hep-ph/0002132}.

\bibitem{Vermaseren:2005qc}
  J.~A.~M.~Vermaseren, A.~Vogt and S.~Moch,
  {\tt hep-ph/0504242}.

\bibitem{Vogt:2000ci}
  A.~Vogt,
  {\it Phys. Lett.} {\bf B 497} (2001) 228,
  {\tt hep-ph/0010146}.

\bibitem{Gardi:2002bk}
  E.~Gardi, G.~P.~Korchemsky, D.~A.~Ross and S.~Tafat,
  {\it Nucl. Phys.} {\bf B 636} (2002) 385,
  {\tt hep-ph/0203161}.

\bibitem{Eynck:2003fn}
  T.~O.~Eynck, E.~Laenen and L.~Magnea,
  {\it JHEP} {\bf 0306} (2003) 057,
  {\tt hep-ph/0305179}.

\bibitem{Anastasiou:2003yy}
  C.~Anastasiou, L.~J.~Dixon, K.~Melnikov and F.~Petriello,
  {\it Phys. Rev. Lett.}  {\bf 91} (2003) 182002,
  {\tt hep-ph/0306192}.

\bibitem{Alekhin20005}
  S. Alekhin, talk presented at the {\it HERA-LHC} Workshop, DESY, 
  March 2005.

\bibitem{Catani:1999hs}
  S.~Catani, M.~L.~Mangano, P.~Nason, C.~Oleari and W.~Vogelsang,
  {\it JHEP} {\bf 9903} (1999) 025,
  {\tt hep-ph/9903436}.

\bibitem{Laenen:2000de}
  E.~Laenen, G.~Sterman and W.~Vogelsang,
  {\it Phys. Rev. Lett.} {\bf 84} (2000) 4296,
  {\tt hep-ph/0002078}.

\bibitem{Sterman:2004yk}
  G.~Sterman and W.~Vogelsang,
  {\it Phys. Rev.} {\bf D 71} (2005) 014013,
  {\tt hep-ph/0409234}.

\bibitem{Kidonakis:1997gm}
  N.~Kidonakis and G.~Sterman,
  {\it Nucl. Phys.} {\bf B 505} (1997) 321,
  {\tt hep-ph/9705234}.

\bibitem{Kidonakis:2000gi}
  N.~Kidonakis and J.~F.~Owens,
  {\it Phys. Rev.} {\bf D 63} (2001) 054019,
  {\tt hep-ph/0007268}.

\bibitem{Dasgupta:2003mk}
  M.~Dasgupta,
  {\it Pramana} {\bf 62} (2004) 675,
  {\tt hep-ph/0304086}.

\bibitem{Kuhlmann:1999sf}
  S.~Kuhlmann {\it et al.},
  {\it Phys. Lett.} {\bf B 476}, (2000) 291,
  {\tt hep-ph/9912283}.

\bibitem{Tzanov:2003gq}
  M.~Tzanov {\it et al.}  [NuTeV Collaboration],
  {\tt hep-ex/0306035}.

\bibitem{Naples:2003ne}
  D.~Naples {\it et al.}  [NuTeV Collaboration],
  {\tt hep-ex/0307005}.

\bibitem{Arneodo:1996qe}
  M.~Arneodo {\it et al.}  [New Muon Collaboration],
  {\it Nucl. Phys.} {\bf B 483} (1997) 3,
  {\tt hep-ph/9610231}.

\bibitem{Benvenuti:1989rh}
  A.~C.~Benvenuti {\it et al.}  [BCDMS Collaboration],
  {\it Phys. Lett.} {\bf B 223} (1989) 485.

\bibitem{Benvenuti:1989fm}
  A.~C.~Benvenuti {\it et al.}  [BCDMS Collaboration],
  {\it Phys. Lett.} {\bf B 237} (1990) 592.

\bibitem{simula}
  M.~Osipenko {\it et al.}  [CLAS Collaboration],
  {\it Phys. Rev.} {\bf D 67} (2003) 092001,
  {\tt hep-ph/0301204}; \\
  M.~Osipenko {\it et al.},
  {\it Phys. Rev.} {\bf D 71} (2005) 054007,
  {\tt hep-ph/0503018}.

\bibitem{Forte:2002fg}
  S.~Forte, L.~Garrido, J.~I.~Latorre and A.~Piccione,
  {\it JHEP} {\bf 0205} (2002) 062,
  {\tt hep-ph/0204232}.

\bibitem{DelDebbio:2004qj}
  L.~Del Debbio, S.~Forte, J.~I.~Latorre, A.~Piccione and J.~Rojo  [NNPDF
  Collaboration],
  {\it JHEP} {\bf 0503} (2005) 080,
  {\tt hep-ph/0501067}.

\bibitem{Forte:2002us}
  S.~Forte, J.~I.~Latorre, L.~Magnea and A.~Piccione,
  {\it Nucl. Phys.} {\bf B 643} (2002) 477,
  {\tt hep-ph/0205286}.

\bibitem{Forte:1998nw}
  S.~Forte and L.~Magnea,
  {\it Phys. Lett.} {\bf B 448} (1999) 295,
  {\tt hep-ph/9812479}.

\bibitem{seligman}
  W.~Seligman, Ph.~D. Thesis, Columbia University, Report 
  No. Nevis-292, 1997.

\bibitem{Catani:1990rr}
  S.~Catani, B.~R.~Webber and G.~Marchesini,
  {\it Nucl. Phys.} {\bf B 349} (1991) 635.

\bibitem{Corcella:2003ib}
  G.~Corcella and A.~D.~Mitov,
  {\it Nucl. Phys.} {\bf B 676} (2004) 346, 
  {\tt hep-ph/0308105}.

\bibitem{Laenen:1998kp}
  E.~Laenen and S.~O.~Moch,
  {\it Phys. Rev.} {\bf D 59} (1999) 034027,
  {\tt hep-ph/9809550}.

\bibitem{Forte:2002ni}
  S.~Forte and G.~Ridolfi,
  {\it Nucl. Phys.} {\bf B 650} (2003) 229,
  {\tt hep-ph/0209154}.


\bibitem{Affolder:2001fa}
  T.~Affolder {\it et al.}  [CDF Collaboration],
  {\it Phys. Rev.} {\bf D 64} (2001) 032001,
  [Erratum-ibid. {\bf D 65} (2002) 039903],
  {\tt hep-ph/0102074}.

\end{thebibliography}
\end{document}